\documentclass[fleqn,10pt]{wlscirep}
\usepackage[utf8]{inputenc}
\usepackage[T1]{fontenc}
\usepackage{float}
\pdfoutput=1
\usepackage{rotating}
\title {DMseg: a Python algorithm for \textit{de novo} detection of differentially or variably methylated regions}

\author[1]{Xiaoyu Wang}
\author[2]{Ming Yu}
\author[1,2,3]{William Grady}
\author[1,5]{Ziding Feng}
\author[1,4,5]{Wei Sun}
\author[1,5,*]{James Y Dai}
\affil[1]{Fred Hutchinson Cancer Center, Public Health Sciences Division, Seattle, 98109, USA}
\affil[2]{Fred Hutchinson Cancer Center, Clinical Research Division, Seattle, 98109, USA}
\affil[3]{University of Washington, Department of Medicine, School of Medicine, Seattle, 98195, USA}
\affil[4]{University of North Carolina, Department of Biostatistics, Chapel Hill, 27516, USA}
\affil[5]{University of Washington, Department of Biostatistics, School of Public Health, Seattle, 98195, USA}

\affil[*]{Correspondence: jdai@fredhutch.org}

\begin{abstract}
Detecting and assessing statistical significance of differentially methylated regions (DMRs) is a fundamental task in methylome association studies. While the average differential methylation in different phenotype groups has been the inferential focus, methylation changes in chromosomal regions may also present as differential variability, i.e., variably methylated regions (VMRs). Testing statistical significance of regional differential methylation is a challenging problem, and existing algorithms do not provide  accurate type I error control for genome-wide DMR or VMR analysis. No algorithm has been publicly available for detecting VMRs. We propose {\tt DMseg}, a Python algorithm with efficient  DMR/VMR detection and significance assessment for array-based methylome data, and compare its performance to {\tt Bumphunter}, a popular existing algorithm.
 Operationally, {\tt DMseg} searches for DMRs or VMRs within CpG clusters that are adaptively determined by both gap distance and correlation between contiguous CpG sites in a microarray. Levene test was implemented for assessing differential variability of individual CpGs. A likelihood ratio statistic is proposed to test for a constant difference within CpGs in a DMR or VMR to summarize the evidence of regional difference. Using a stratified permutation scheme and pooling null distributions of LRTs from clusters with similar numbers of CpGs, {\tt DMseg} provides accurate control of the type I error rate. In simulation experiments, {\tt DMseg} shows superior power than {\tt Bumphunter} to detect DMRs. Application to methylome data of Barrett's esophagus and esophageal adenocarcinoma reveals a number of DMRs and VMRs of biological interest. \\

{\tt DMseg} is available at   {\tt https://pypi.org/project/DMseg}
\end{abstract}

\begin{document}

\flushbottom
\maketitle
%
%
\thispagestyle{empty}

\section*{Introduction}
Epigenetics, DNA modifications that do not change the underlying sequence, may provide an interface between environmental insult, genetic susceptibility and disease development. There is increasing interest to study epigenetics and understand the etiology of complex diseases such as cancers \cite{Petronis2010,Esteller2008}. The most common epigenetic markers are
CpG methylation, a chemical modification of a cytosine (methyl-cytosine) that is immediately followed by a guanine. With advances in high-throughput technologies, it is now possible to measure DNA methylation for nearly all 28 million CpGs in the human genome by whole-genome bisulfite squencing (WGBS), or more economically, a highly selected subset with direct functional relevance measured by microarrays such as Illumina Infinium HumanMethylation450 and EPIC BeadChips. Methylome association studies are now routinely conducted to understand etiology or identify disease biomarkers \cite{Relton2010}.

In human genome, CpG methylation sites are irregularly spaced, with blocks of correlated CpGs, e.g., promoters, CpG island and CpG island shores. Consequently, when comparing samples under different biological conditions, differential methylation sites were often found in regions with contiguous CpGs simultanuously hyper- or hypo- methylated \cite{Irizarry2009}. The identification and characterisation of differentially methylated regions (DMRs) between phenotype groups have been of prime interest in methylome association studies \cite{Jaffe2012a}. For cancer methylome, differential methylation can often present as mean methylation shift as well as increased stochastic variance, the latter of which is due to the heterogeneous nature of cancers \cite{Hansen2011,Phipson2014}. The main methodological challenge is to discover differentially methylated region (DMR) or variably methylated region (VMR) in an efficient manner, and to provide an accurate assessment of statistical significance of these regional findings.

There are a number of DMR calling algorithms for the restricted CpG coverage in HM450 arrays and EPIC BeadChips \cite{Jaffe2012a,Pedersen2012,Peters2015}.  Nearly all existing DMR calling methods start from computing individual CpG associations and identifying contiguous ones with evidence of differential methylation. Local smoothing of probe-level statistics with respect to chromosomal coordinates is commonly adopted, for example, {\tt Bumphunter} and {\tt DMRcate}. Except for {\tt Bumphunter}, rigorous evaluation of statistical significance for DMR findings has not always conducted due to computational burden required by a permutation or bootstrap procedure. Indeed, in our data applications it may take several hours for {\tt Bumphunter} to compute family-wise error rate for discovered DMRs. On the other hand, methods for detecting VMR have been proposed \cite{Jaffe2012}, though computational software for genome-wide VMR are not yet readily available.

In this article, we develop a unified method for detecting and assessing {\sl de novo} DMR/VMR in methylome association studies. A computationally efficient Python algorithm, {\tt DMseg}, is released at PyPi.org to implement the proposed method. Comparing to existing methods and algorithms, {\tt DMseg} made several contributions. First, a likelihood ratio test (LRT) statistic assuming a common difference across contiguous CpGs is used to summarize the evidence of DMR/VMR. This test provides some level of smoothing by averaging CpG-level associations, though much simpler than local smoothing based on chromosomal coordinates. Second, we show that test statistics for DMRs are typically not exchageable between DMRs with different numbers of CpGs. To unbiasely assess the false positive error rate, a stratified permutation test was implemented in {\tt DMseg} that accounts for non-exchangeability of LRT statistic. We conduct extensive simulations to evaluate and compare  {\tt DMseg} to {\tt bumphunter}, the only existing algorithm with significance assessment for DMR.

\section*{ Methods \& Datasets}

The objective of {\tt DMseg} is to enable fast detection of DMR and VMR for microarray-based methylome association studies, and to provide an accurate assessment of statistical significance of the resultant DMRs and VMRs. Many algorithms exist for detecting DMR \cite{Jaffe2012a,Pedersen2012,Peters2015}, though few is capable of conducting a rigorous hypothesis test for DMR findings.

We use methylome studies on progression of a cancer precursor lesion, Barrett's esophagus, to esophageal adenocarcinoma to illustrate the motivation and the utility of {\tt DMseg}. The scientific goal is to discover and validate early detection methylation biomarkers for malignant transformation of BE, so that preventative measures can be undertaken to intercept cancer progression. Two methylome datasets were downloaded from GEO: the first dataset contains 64 normal squamous esophagus samples, 19 Barrett$\prime$s esophagus (BE) and 125 esophageal adenocarcinoma (EAC) samples from Australia \cite{Krause2016}. After quality control, there were 372376 CpGs for analysis; the second set is from 33 normal squamous samples, 59 BE samples from cancer free patients, and 23 EAC samples from United States \cite{Yu2019}. All samples were fresh frozen and profiled by Infinium Human Methylation 450K BeadChips.

\subsection*{Algorithm modules and workflow}
The input datasets for  {\tt DMseg} include an epidemiologic dataset containing the primary comparison groups for association and adjusting covariates, and a methylation data matrix for processed methylome data. Suppose there are $n$ samples with DNA methylome data available for the association analysis, denoted by $(\mbox{\bf Y}_i,X_i,W_i)$, for $i=1,...,n$, where $\mbox{\bf Y}_i$ is a length $p$ vector for methylation beta values or M-values in $p$ CpG sites, $X_i$ is the group indicator of the primary interest,  for example the cancer ($X_i=1$) or normal sample ($X_i=0$), and $W_i$ is the vector of additional covariates such as age and gender that should be adjusted for. Let $Y_{ij}$ denote the beta value or M-value of the $j^{th}$ CpG site for the $i^{th}$ sample.

\begin{figure}[H]
\centerline{\includegraphics[width=200pt]{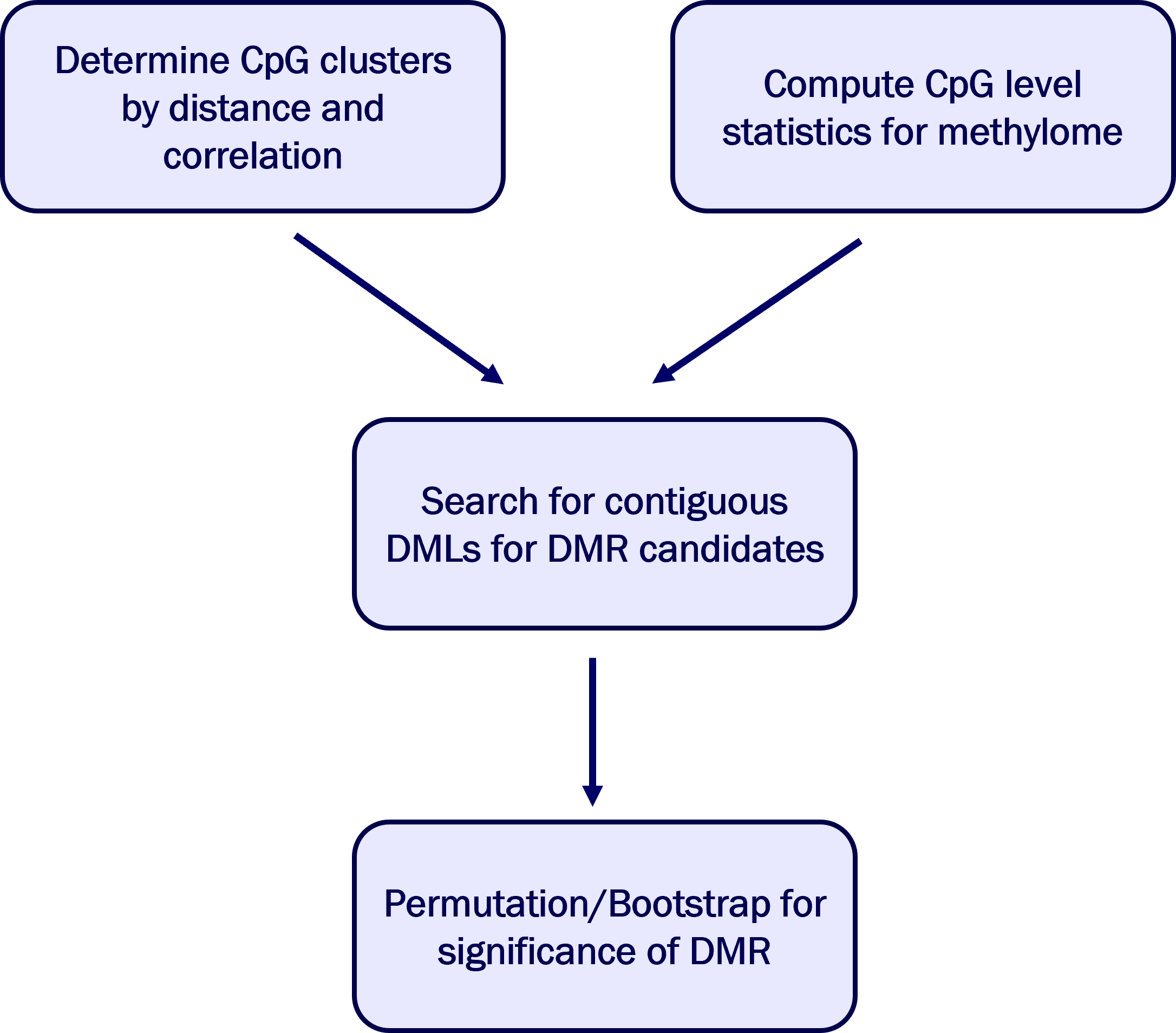}}
\caption{The modules and the workflow of {\tt DMseg} algorithm}\label{fig:1}
\end{figure}

Algorithmically, {\tt DMseg} is composed of four modules (Figure~1\vphantom{\ref{fig:1}}): determine clusters of CpGs by distance and correlation; compute CpG-level association statistics; search clusters for candidate DMRs and computing likelihood ratio statistics; assess significance of DMRs by a permutation or bootstrap procedure.

\noindent {\bf Determining clusters of CpGs: } The CpG sites selected for the HM450 array and the EPIC BeadChip represent a small subset (1.5-3\%) of the entire 28 million CpG sites in the human genome, mostly distributed in the CpG-rich regions such as promoters, CpG islands and shores. The spacing between CpGs are highly irregular: clusters of densely distributed, highly correlated CpGs are separated by long intergenic regions with scarce CpGs. Allowing parallel searching, a computationally efficient approach is to first identify clusters and then detect DMRs within clusters, as implemented in {\tt bumphunter}. {\tt DMseg} determines the two contiguous CpGs belong to a cluster if the distance between them is less than a user defined number of base pairs (e.g. 500), or the correlation between contiguous CpGs greater than 0.6 (the median correlation coefficent among contiguous CpGs within the clusters defined by gap distance only). This criterion is different from {\tt bumphunter} \cite{Jaffe2012a}, which only uses the gap distance to determine clusters. The motivation is that correlation may not always decay linearly with chromosomal distance in base pairs, and there are CpGs far away in chromosomal coordinates yet highly correlated.

Using the HM450 data for BE/EAC samples as an example, we obtained 170156 clusters (including 1-CpG clusters) by the maximum gap distance 500 bp. Adding the correlation criterion will reduce the cluster number to 150676, of which 19480 clusters were connected due to correlation. Removing 1-CpG clusters will result in 54688 clusters, from which DMRs were searched for.

\noindent {\bf Computing CpG-level association statistics: } Users can choose whether to use methylation beta values or M values for calculating individiual CpG association for testing differentially methylated CpGs. The former yields regression coefficents are directly interpretable as the average percentage of methylation change between comparison groups, while the latter works better for statistical performance of linear models. Let $Y_{ij}$ denote the methylation value for the $i^{th}$ sample and the $j^{th}$ CpG site. For testing differentially methylated CpGs, the following linear model is fitted for each of CpGs in the array,
\begin{eqnarray}
  \mathbb{E}(Y_{ij})&=& \beta_{0j} + \beta_{1j} X_i + \beta_{2j} W_i,  \label{eq:model1}
\end{eqnarray}
where $\beta_{1j}$ is the association parameter of interest. For testing variably methylated CpGs, $Y_{ij}$ is the methylation M-value and the following regression model will be fitted
\begin{eqnarray}
  \mathbb{E}(|Y_{ij} - \widetilde{m}_{g(i)j}|)&=& \alpha_{0j} + \alpha_{1j} X_i + \alpha_{2j}W_i,  \label{eq:model2}
\end{eqnarray}
where $g(i)$ is the group (e.g., tumor or normal) label for $i^{th}$ sample, $\widetilde{m}_{g(i)j}$ is the sample median for $j^{th}$ CpG site in $g(i)$ group, $|Y_{ij} - \widetilde{m}_{g(i)j}|$ is the absolute difference to the corresponding group median. Model (\ref{eq:model2}) implements the Levene test for differential variability \cite{Brown1974}. Because the predictors and the design matrix for each of CpG association models are identical,  the estimated $\beta_{1j}$s and its estimated standard errors can be computed by sharing common intermediary elements of ordinary least squares regression for different CpGs. Therefore the algorithm does not need to iterate through CpGs one by one, which substantially reduces the computation time.

\noindent {\bf Detecting DMR/VMR and computing likelihood ratio statistics: } We consider candidate DMR/VMR to be regions with at least $k$ contiguous  CpGs that are differentially methylated, where $k=2$ or 3 as defined by users. Operationally, searching for candidate DMR/VMR is conducted within clusters that are previously determined, using $z$-score for individual CpG association $\geq$ 1.96 as a soft threshold for differential methylation. Two features of {\tt DMseg} permit flexibility in accommodating diverse genomic irregularities: first, {\tt DMseg} allows switching sign of CpG association (hypermethylation or hypomethylation) within a DMR, similar to {\tt DMRcate} \cite{Peters2015}, since sometimes hypermethylation in gene bodies may also be associated with upregulation of gene expressions; second, though {\tt DMseg} don't employ any smoothing method for CpG level association statistics, if two adjacent DMRs or two adjacent VMRs within one cluster are separated by merely a single CpG, and that CpG shows a moderate level of association, e.g., $z$-score $\geq$ 1.64 (user can define this softer threshold), then the two DMRs or VMRs will be merged through the CpG.

After a candidate DMR was detected, a likelihood ratio statistic will be computed for the DMR, comparing the null model that there is no association for any CpGs within the DMR, to the alternative model that all CpG-level association parameters within a DMR are a same scalar. Suppose $\widehat{\mbox{\boldmath $ \beta $}}$ =$(\hat{\beta}_{11},...,\hat{\beta}_{1L})$ are estimated association parameters for $L$ CpGs within a candidate DMR. Let $\hat{\sigma}^2_{11},...,\hat{\sigma}^2_{1L}$ denote the variance of $\hat{\beta}_{11},...,\hat{\beta}_{1L}$ respectively. When sample size $n$ is sufficiently large, the distribution of $\hat{\sigma}^2_{11},...,\hat{\sigma}^2_{1L}$ can be approximated by a Gaussian distribution with its corresponding variance. Under the alternative hypothesis that all $\beta_{1j}$ are the same constant $\bar{\beta}_{1}$, the maximum likelihood estimate of $\bar{\beta}_{1}$ is $\mbox{\bf J}'\Sigma^{-1} \widehat{\mbox{\boldmath $ \beta $}} $/$\mbox{\bf J}'\Sigma^{-1} \mbox{\bf J} $, where $\mbox{\bf J}$ is the length-$L$ vector of 1, and $\Sigma$ is $L \times L$ diagonal matrix with its diagonal elements $\hat{\sigma}^2_{11},...,\hat{\sigma}^2_{1L}$. The LRT statistic is therefore
\begin{equation}
\widehat{\mbox{\boldmath $ \beta $}}'\Sigma^{-1} \widehat{\mbox{\boldmath $ \beta $}} - (\widehat{\mbox{\boldmath $ \beta $}} - \bar{\beta}_{1}) '\Sigma^{-1} (\widehat{\mbox{\boldmath $ \beta $}} - \bar{\beta}_{1}).
\end{equation}
The larger this LRT is, the greater evidence for a significant DMR. As we show next, a larger LRT could be resulted from a larger difference in mean methylations, or a DMR with more CpGs included.

\begin{figure}[H]
\centerline{\includegraphics[width=300pt]{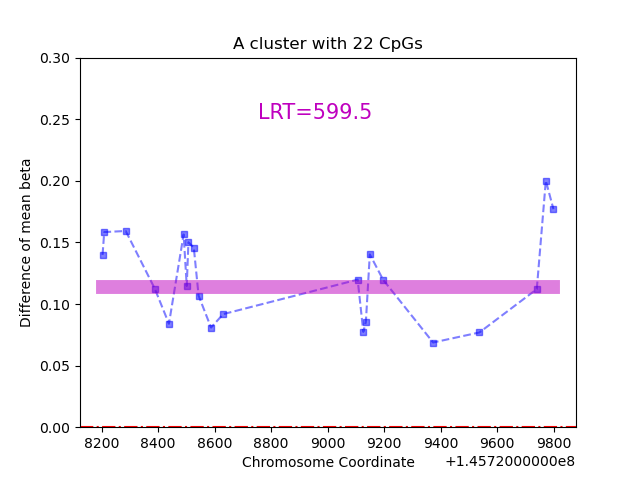}}
\caption{An example of DMR fitted by {\tt DMseg} algorithm}\label{fig:2}
\end{figure}

Figure~2\vphantom{\ref{fig:2}} shows an example DMR in chromosome 8. There are 22 CpGs within this cluster (starting position 145728203, ending position 145729799), all of which showed hypermethylation in esophageal adenocarcinoma samples relative to Barrett's esophagus samples. The z-scores of all 22 CpG associations are greater than 2.0, suggesting a highly significant DMR.  {\tt DMseg} fitted a horizontal line to the 22 differences of beta values with its height slightly greater than 0.1, and the LRT for this DMR is 599.5.  \\

\begin{figure}[H]
\centerline{\includegraphics[width=450pt]{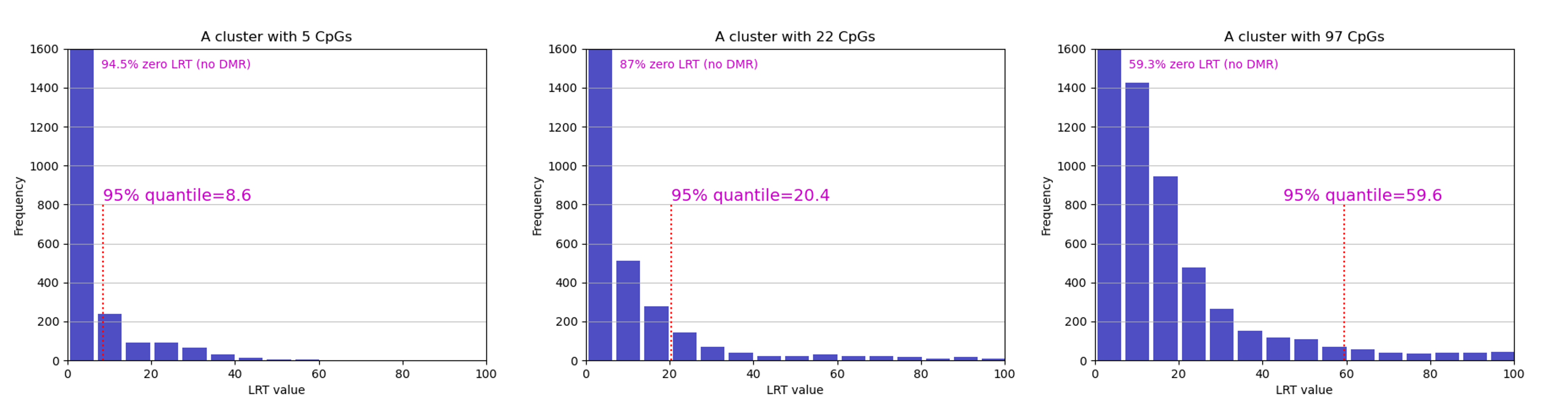}}
\caption{Clusters with different number of CpGs have different null distributions for LRT.}\label{fig:3}
\end{figure}

\noindent {\bf Significance test of DMRs by permutation or bootstrap: }  One distinguishing feature of {\tt DMseg} is to conduct a permutation/bootstrap test for assessing the significance of DMR. This can be computationally demanding for genome-wide testing. For example, it takes several hours for {\tt bumphunter} to conduct its bootstrap-based testing procedure. Moreover, hypothesis testing for a data-driven DMR finding can be challenging because: i) it is difficult to establish the null distribution of such data-driven DMR; ii) the test statistic, e.g., LRT for {\tt DMseg}  may not be ``exchangeable'' for different clusters. Figure~3\vphantom{\ref{fig:3}} presents the permutation null distributions from three clusters with different sizes (5 CpGs, 22 CpGs, 97 CpGs). Clearly clusters with a greater number of CpGs have larger LRT values:  94.5\% of permutations from the 5-CpG cluster yield no DMR finding (therefore LRT=0), and the 95\% quantile of its null distribution (including LRT=0 permutations) is 8.6; 87\% of permutations from the 22-CpG cluster yield no DMR finding, and the 95\% quantile of its null distribution  is 20.4; 59.3\% of permutations from the 97-CpG cluster yield no DMR finding, and the 95\% quantile of its null distribution is 59.6. In summary, more CpGs in the cluster to start the searching lead to a higher probability of finding DMR, and a DMR with more CpGs yields a larger LRT.

This ``non-exchageability'' property for LRT statistics suggests that one cannot pool null LRT statistics across clusters with different numbers of CpGs. Ideally, a large number of permutations need to be conducted for each cluster to result in a p-value with $10^{-8}$ resolution, which is computationally prohibitive. {\tt DMseg} implements a stratified pooling strategy: pooling permutation LRTs as the null distribution for computing p-values from 4 strata with CpG numbers in (0,10],(10,20],(30,40] and more than 40. For clusters without DMR findings, their p-values are set to be 1. FWERs for top DMR candidates are then calculated from these p-values.

\section*{Results}
\subsection*{Synthetic data to evaluate type I error rate and power}
A simulation experiment was conducted to evaluate the type I error rate and the power of {\tt DMseg} and {\tt Bumphuner}. To mimic real methylome data, the synthetic data were created using on the normal DNA methylation data in the BE/EAC example \cite{Krause2016} as the backbone, and adding a gradient of effect sizes for differential methylation. Specifically, the data for the null hypothesis were constructed by randomly splitting the 64 normal squamous esophagus samples to two groups for 100 times, each with 32 samples. To reduce the computation time, we restricted to all CpGs data in chromosome 10, and supplemented CpGs from clusters with numbers of CpGs $>$ 20 from all other chromosomes. The rationale is to test the performance in the full spectrum of CpG clusters, particularly large clusters with many CpGs. In total, there are 29165 CpGs included, and 9975 CpGs of them are from other chromosomes. These CpGs form 3197 clusters.

\begin{figure}[H]
\centerline{\includegraphics[width=300pt]{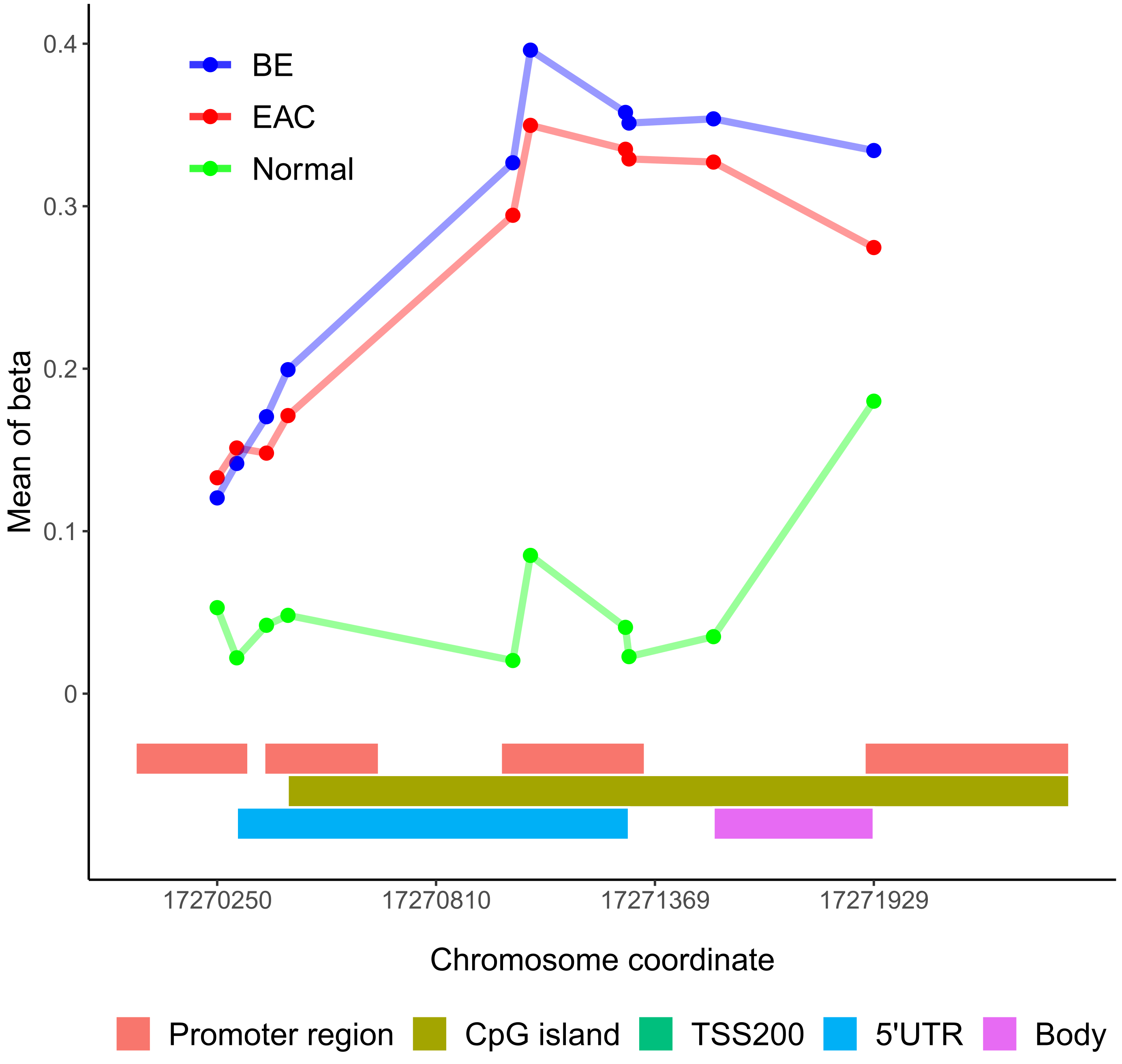}}
\caption{The mean CpG methylation levels for the ten CpGs in VIM gene chosen for the simulation study.}\label{fig:4}
\end{figure}

To generate a DMR and test the power, we selected 10 CpGs in the \emph{Vimentin} gene and added signals in either the beta value scale or the M value scale to every sample in the designated ``case'' group. \emph{Vimentin} gene in Chromosome 10 is a known methylation biomarker for detecting BE and EAC \cite{Moinova2018}. Figure~4\vphantom{\ref{fig:4}} shows the mean beta values among 10 CpGs (from cg05151811 to cg02746869) in \emph{VIM} gene for BE, EAC, and normal samples. There are marked differences in mean beta values between normal samples and BE/EAC samples for the 10 CpGs in the promoter region, though the differences between BE and EAC are less pronounced. As shown in Table \ref{tenCpG_simulations}, to evaluate the full-range power performance four levels of DMR effect sizes were added to the beta values of the 10 CpGs for every sample in the ``case'' group, and similarly five levels of effect sizes in the M-value scale. Note that when effect size is set to be zero, the family wise error rate (FWER, the probability of declaring one or more false positives) was evaluated.

The comparator {\tt Bumphunter} can be used with customized options. For example, users can pick a cutoff value for the minimal differences of mean beta values or mean M values for contiguous CpGs to be considered as DMR, or users can let {\tt Bumphunter} decide the cutoff value by simulations. While {\tt Bumphunter} proposed to use smoothed CpG methylation differences for detecting DMR, users can choose the option of no smoothing and therefore using the original methylation differences. The results for  {\tt Bumphunter} under these different options were shown in Table \ref{tenCpG_simulations}. In either scale, {\tt DMseg} preserved the correct FWER in 100 permutation datasets, and delivered a much improved power over {\tt Bumphunter}, regardless of the options used in {\tt Bumphunter}. When {\tt Bumphunter} was set to automatically select a cut-off value for detecting DMR, FWER is inflated whether smoothing was used or not. The smoothing option clearly worsened the performance of {\tt Bumphunter}

Table \ref{fiveCpG_simulations} shows the results when only the last five CpGs were used in the simulation experiment. This scenario was created to evaluate power in weaker signal-to-noise settings, and may present challenges to the smoothing option of {\tt Bumphunter}. The superior performance of {\tt DMseg} over {\tt Bumphunter} is more drastic for this scenario with 5 differentially methylated CpGs.

\begin{table}[H]
\centering
\caption{Simulation results for type I error rate and power when adding differential methylation to 10 CpGs in VIM gene \label{tenCpG_simulations}}
\begin{tabular}{llllll}
\toprule \multicolumn{2}{l}{{\begin{tabular}[c]{@{}l@{}}Analysis \\ scale\end{tabular}}} & {\begin{tabular}[c]{@{}l@{}}Effect size in \\ chosen scale\end{tabular}} & \multicolumn{2}{l}{\begin{tabular}[c]{@{}l@{}}Bumphunter No smoothing\\ (smoothing)\end{tabular}} & {DMseg} \\  \cline{4-5}
\multicolumn{2}{l}{}    &   & Set threshold\textsuperscript{a}  & \begin{tabular}[c]{@{}l@{}}Automatically \\pick threshold\end{tabular}      &   \\ \midrule
{Beta} & {FWER}   & 0  & 0.03 (0.03) & 0.08 (0.11) & 0.05  \\ \cline{2-6}
  & {Power}  & 0.05  & 0 (0) & 0 (0) & 0.53 \\ \cline{3-6}
  &          & 0.1  & 0.07 (0)  & 0 (0)  & 0.97 \\ \cline{3-6}
  &          & 0.15 & 0.86 (0.72) & 0.05 (0)  & 1  \\ \midrule
{M-value}   & FWER  & 0  & 0.05 (0.06) & 0.06 (0.07) & 0.06 \\ \cline{2-6}
  & {Power}  & 1  & 0 (0)  & 0.07 (0)  & 0.48  \\ \cline{3-6}
  &         & 1.2  & 0 (0)  & 0.29 (0)  & 0.74  \\ \cline{3-6}
  &         & 1.4  & 0 (0)  & 0.54 (0)  & 0.94  \\ \cline{3-6}
  &         & 1.6  & 0 (0)  & 0.72 (0.01)  & 0.94 \\ \hline
\end{tabular}
\footnotesize{\\$^a$ 0.1 used for beta values, 0.28 for M-value}\\
\end{table}

\begin{table}[H]
\centering
\caption{Simulation results for type I error rate and power when adding differential methylation to 5 CpGs in VIM gene \label{fiveCpG_simulations}}
{\begin{tabular}{llllll}
\toprule \multicolumn{2}{l}{{\begin{tabular}[c]{@{}l@{}}Analysis \\ scale\end{tabular}}} & {\begin{tabular}[c]{@{}l@{}}Effect size in \\ chosen scale\end{tabular}} & \multicolumn{2}{l}{\begin{tabular}[c]{@{}l@{}}Bumphunter No smoothing\\ (smoothing)\end{tabular}} & {DMseg} \\  \cline{4-5}
\multicolumn{2}{l}{}    &   & Set threshold\textsuperscript{a}  & \begin{tabular}[c]{@{}l@{}}Automatically \\pick threshold\end{tabular}      &   \\ \midrule
{Beta} & {FWER}   & 0  & 0.03 (0.06) & 0.08 (0.11) & 0.05  \\ \cline{2-6}
  & {Power}  & 0.1  & 0 (0) & 0 (0) & 0.85 \\ \cline{3-6}
  &          & 0.15  & 0.07 (0)  & 0 (0)  & 0.98 \\ \cline{3-6}
  &          & 0.2  & 0.01 (0.02) & 0 (0)  & 1  \\ \midrule
{M-value}   & FWER  & 0  & 0.05 (0.06) & 0.06 (0.08) & 0.06 \\ \cline{2-6}
  & {Power}  & 1.6  & 0 (0)  & 0 (0)  & 0.44  \\ \cline{3-6}
  &         & 2  & 0 (0)  & 0.06 (0)  & 0.75  \\ \cline{3-6}
  &         & 2.4  & 0 (0)  & 0.33 (0)  & 0.95  \\ \cline{3-6}
  &         & 2.8  & 0 (0)  & 0.55 (0.01)  & 0.97  \\ \hline
\end{tabular}}
\footnotesize{\\$^a$ 0.1 used for beta values, 0.28 for M-value}\\
\end{table}

\subsection*{HM450 data example: methylome comparison between BE and EAC}
{\tt DMseg} was used in a biomarker study to identify DMR and VMR between BE and EAC. The Australian dataset was used as the discovery set \cite{Krause2016}, and the US dataset was used as the validation set \cite{Yu2019}. As a comparison, {\tt Bumphunter} did not yield any significant DMR at FWER 0.05 level.


Using {\tt DMseg}, we identified 47 significant DMRs in 16 chromosomes and 2 significant VMRs in 2 chromosomes after correcting for multiple testing (FWER<0.05). A linux machine with Intel(R) Xeon(R) Gold 6254 CPU (3.1 GHz), and 64 GB memory was used in this study. The most recent version of {\tt Bumphunter}) (V1.38.0) was run on R (V4.2.0), and {\tt DMseg} was run on Python (V3.8.12). It took {\tt DMseg} 12 minutes 10 seconds to run DMR analysis for this discovery set (23 minutes 33 seconds if using {\tt Bumphunter}). The top 10 DMRs and the two VMRs are shown in Table \ref{DMRVMR}, ordered by FWER. The full list of DMRs are included in Supplementary Table 1. For each local cluster that contains a DMR or VMR identified in the discovery set, we ran {\tt DMseg} in the validation set see if the same or similar DMR can be detected in the cluster. Sometimes not all CpGs in a DMR were available for validation, due to removal of CpGs during the quality control steps of the two datasets. Strikingly, all DMRs and VMRs identified in the discovery set were validated with a p-value less than 0.05 (Table \ref{DMRVMR}, Supplementary Table 1). A high degree of concordance was observed for LRTs and segment means for DMRs in the discovery set and in the validation set, supporting the robustness of these DMR findings.

\begin{figure}[H]
\centerline{\includegraphics[width=500pt]{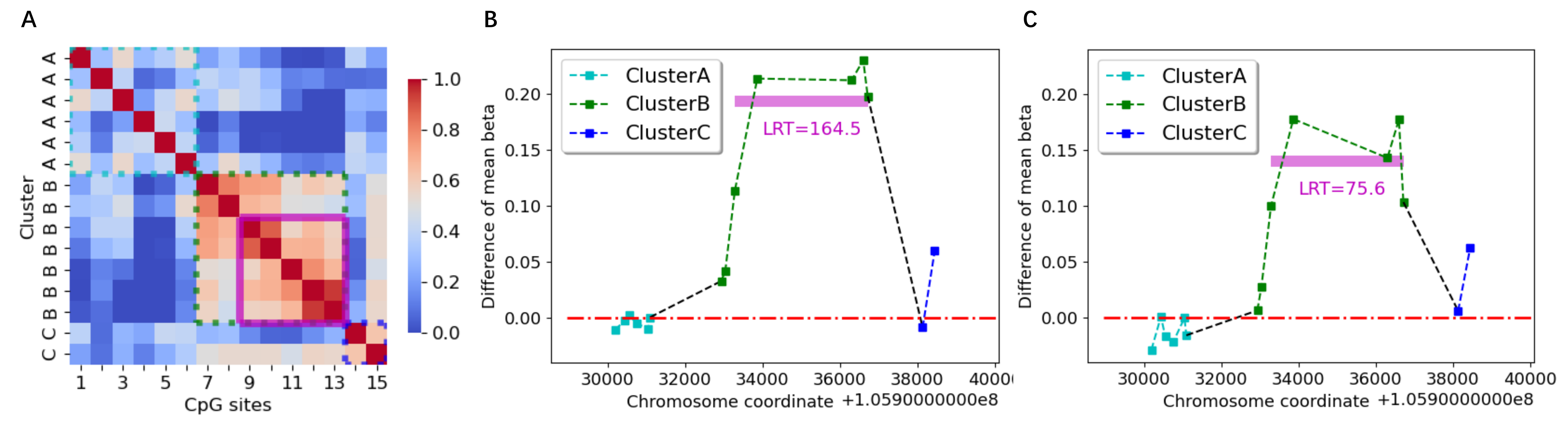}}
\caption{A DMR detected by DMseg in the Discovery set. (A) Pearson correlation among CpGs in Clusters A-C. Dashed boxes in different colors show correlation among CpGs in different clusters, and the solid purple box indicates correlation among CpGs within the DMR. (B) DMR (the purple box) detected in this region. (C) The same DMR in the Validation set.}\label{fig:5}
\end{figure}

Figure~5\vphantom{\ref{fig:5}} shows a significant DMR in Chromosome 14 detected by {\tt DMseg}, ranked the 39$^{th}$ among the 47 DMRs. This is an example in which merging long separated CpGs by correlation led to new discoveries. This is a region in \emph{MTA1} gene, which have been reported to promote tumorigenesis and development of esophageal squamous cell carcinoma, also a potential indicator for assessing the malignant potential of colorectal and gastric carcinomas \cite{Nan2020,Toh1997}. The correlation matrix of CpGs in this region is shown by a heatmap in the left panel; The difference of mean beta along chromosome coordinates are shown in the right panels  (CpGs in different clusters are indicated by different colors). The Cluster B is an example of connecting contiguous CpGs with high correlation but a longer distance than the max gap 500 bp: without merging by correlation, there would be 3 clusters in Cluster B with the sizes of 3 CpGs, 1 CpG, 3 CpGs respectively. Connecting these CpGs created a larger cluster, which led to a statistically significant DMR.

\begin{figure}[H]
\centerline{\includegraphics[width=500pt]{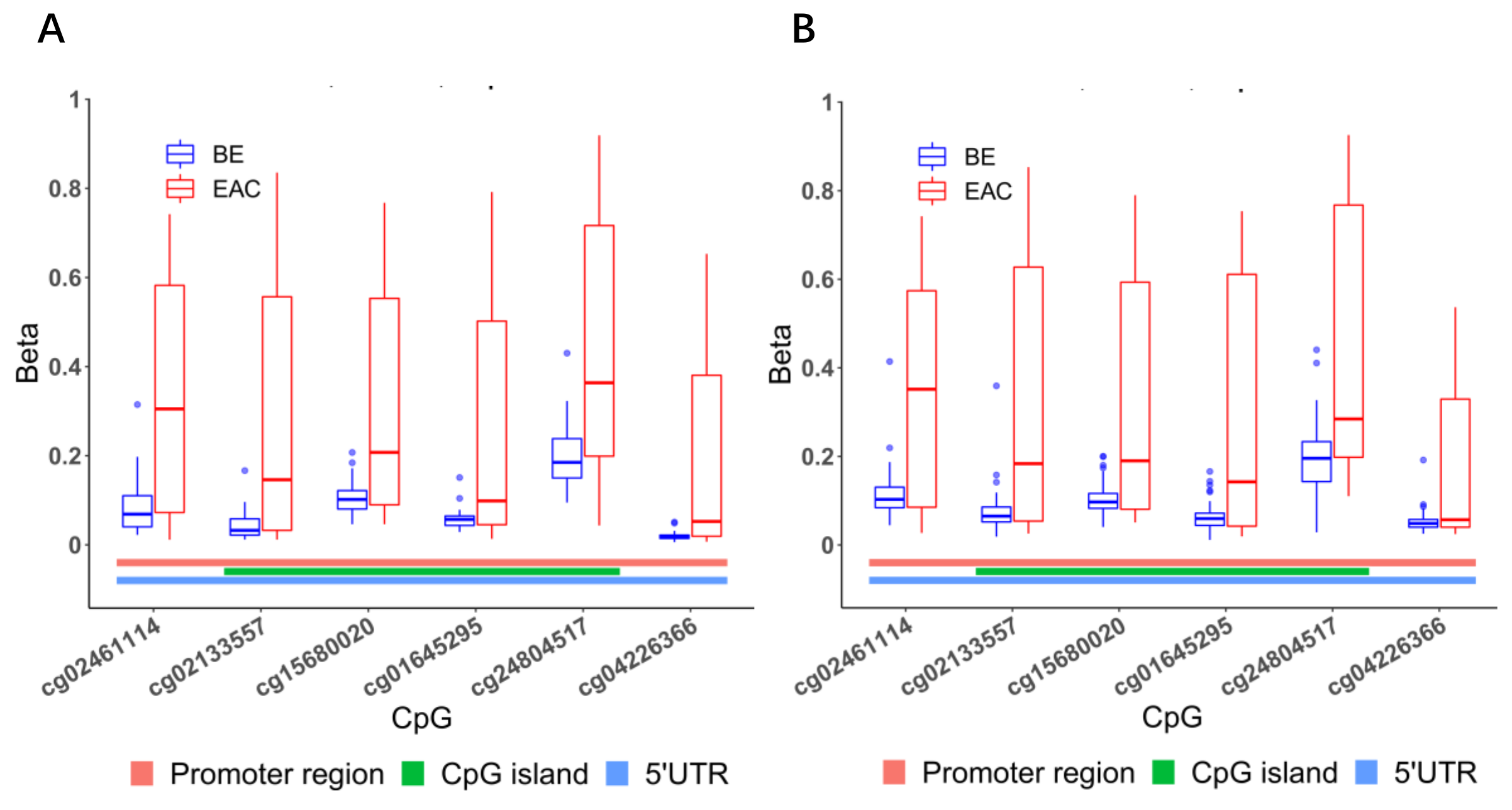}}
\caption{A VMR detected by DMseg. (A) VMR in the Discovery set. (B) The same VMR in the Validation set.}\label{fig:6}
\end{figure}

The two significant VMRs in Table \ref{DMRVMR} are not overlapped with any of the 47 DMRs, though they also have a  moderate level of mean differences between BE and EAC samples (p-values for testing DMR are $2.52\times10^{-5}$ and $7.37\times10^{-4}$, with corresponding FWER 0.496 and 1). For example, the top ranked VMR (Figure~6\vphantom{\ref{fig:6}}) contains 6 CpGs in Chromosome 20 in the promoter region of \emph{DIDO1} gene, which has been reported to play an important role in inhibiting gastric cancer progression \cite{Zhang2021}. All six CpGs showed low methylation in the BE samples, and significantly increased variability and hypermethylation (though did not reach 0.05 in FWER) in EAC samples. The mean standard deviation of beta values for 6 CpGs in BE samples is 0.046 (mean methylation level 0.09), and it increases to 0.25 for EAC samples (mean methylation level 0.31). The statistical significance for the increased variability is visibly stronger than that for hypermethylation.

\begin{sidewaystable}
\caption{Differentially methylated regions in the BE-EAC comparison}
\label{DMRVMR}
\begin{tabular}{lllccccccc|cccc}
\toprule
{}  & {chr} & {gene} & \multicolumn{7}{c}{Discovery set} & \multicolumn{4}{c}{Validation set}   \\ \cline{4-14}
&   &  & start CpG  & end CpG   & \#CpG & mean  & LRT   & p-value      & FWER   & \#CpG & mean & LRT   & p-value \\ \hline
{DMR} & chr2  & \emph{CAPN10}  & cg08589214 & cg16615776 & 5  & 0.22  & 599.5 & \textless{}$1\times10^{-6}$ & \textless{}0.002 & 5 & 0.21  & 326.4 & \textless{}0.002 \\ \cline{2-14}
      & chr11  & \emph{AP2A2;MUC6}  & cg21625737 & cg20839554 & 6 & 0.19  & 409.4 & \textless{}$1\times10^{-6}$ & \textless{}0.002 & 6  & 0.20  & 275.5 & \textless{}0.002 \\ \cline{2-14}
      & chr3  & \emph{EPHB3}  & cg03867377 & cg10769844 & 5  & 0.15 & 352.5 & \textless{}$1\times10^{-6}$ & \textless{}0.002 & 5     & 0.14       & 124.7 & \textless{}0.002 \\ \cline{2-14}
      & chr11  & \emph{NUMA1}  & cg02151120 & cg05874355 & 6  & 0.22 & 345.0 & \textless{}$1\times10^{-6}$ & \textless{}0.002 & 7  & -0.05;0.22 & 329.2 & \textless{}0.002 \\ \cline{2-14}
      & chr1  & \emph{SNORD103A/PUM1}  & cg24722354 & cg00581541 & 4  & 0.22 & 311.7 & \textless{}$1\times10^{-6}$ & \textless{}0.002 & 3  & 0.17  & 66.6  & \textless{}0.002 \\ \cline{2-14}
      & chr13  & \emph{A2LD1}  & cg02425595 & cg25710107 & 9  & 0.14 & 303.6 & \textless{}$1\times10^{-6}$ & \textless{}0.002 & 9  & 0.11  & 155.2 & \textless{}0.002 \\ \cline{2-14}
      & chr4  & \emph{LOC100130872/SPON2}  & cg20756245 & cg17232217 & 16  & 0.13 & 293.9 & \textless{}$1\times10^{-6}$ & \textless{}0.002 & 17  & 0.16  & 507.4 & \textless{}0.002 \\ \cline{2-14}
      & chr1  & \emph{GJC2;GUK1}  & cg22844623 & cg20848130 & 9  & 0.16  & 276.0 & \textless{}$1\times10^{-6}$ & \textless{}0.002 & 9  & 0.17  & 258.4 & \textless{}0.002 \\ \cline{2-14}
      & chr14  & \emph{ZFYVE21}  & cg13154413 & cg25580656 & 6  & 0.24 & 252.1 & \textless{}$1\times10^{-6}$ & \textless{}0.002 & 6  & 0.17  & 169.5 & \textless{}0.002 \\ \cline{2-14}
      & chr14  & \emph{JDP2}  & cg25102206 & cg22143352 & 4  & 0.24 & 247.5 & \textless{}$1\times10^{-6}$ & \textless{}0.002 & 4  & 0.23  & 322.8 & \textless{}0.002 \\ \hline
{VMR} & chr20  & \emph{DIDO1}  & cg02461114 & cg04226366 & 6  & 1.38  & 179.4 & $4.39\times10^{-8}$  & 0.002  & 6  & 1.48  & 432.8 & \textless{}0.002 \\ \cline{2-14}
      & chr19  & \emph{ZNF154} & cg03142586 & cg27324426 & 10 & -0.72 & 164.3 & $1.32\times10^{-7}$ & 0.006  & 4  & -0.43 & 20.5  & 0.032  \\ \hline
\end{tabular}
\end{sidewaystable}

\newpage
\subsection*{Epic array data application}

Other than the HM450 array, {\tt DMseg} can analyze data generated by the Infinium Methylation EPIC microarray, which provides a higher throughput platform to quantify methylation at over 850,000 CpG sites on the human genome. To test speed, we analyzed a Fred Hutch dataset containing 11 Gastroesophageal reflux disease (GERD) samples and 10 BE samples. This data include 760241 CpGs after quality control steps. Total number of clusters (including 1 CpG clusters) is 388459 before merging, 367483 after merging. These numbers are about double of those for BE and EAC data on the HM450 platform. It took 16 minutes 41 seconds for {\tt DMseg} to run DMR analysis (34 minutes 4 seconds for {\tt Bumphunter}). Due to the small sample size, no DMR was found with FWER $<$ 0.05. On the other hand, one VMR with 7 CpGs (from cg05087455 to cg19240637 in Chromosome 2) was found in \emph{RNF144A} gene with  LRT = 97.2 and FWER = 0.046.

\section*{Discussion}

In this work we proposed a fast and accurate algorithm for detecting statistically significant DMR/VMR for array-based methylome data, {\tt DMseg}. The superior performance of {\tt DMseg} can be attributed to several features of {\tt DMseg}: 1) {\tt DMseg} uses Z-score statistics to detect contiguous DMR/VMR, which is statistically principled and more sensitive than the differential methylation regression coefficients used in {\tt Bumphunter}; 2) {\tt DMseg} uses a likelihood ratio test statistic that fits a constant mean to a DMR, averaging differential methylation effect sizes among CpGs in a DMR. This avoids commonly used smoothing procedures (loess or kernel smoothing) that requires bandwidth selection. The latter is quite challenging for irregularity  and sparse distribution of CpG coverage in a methylation array; 3) {\tt DMseg} defines clusters within which searching for DMR is conducted using both gap distance and correlation, which is more data-adaptive than {\tt Bumphunter}. This creates longer clusters and yields more DMRs; 4) {\tt DMseg} pools null distributions of LRT for clusters with similar numbers of CpGs, effectively increasing the power for detecting shorter DMRs.

The most computationally expensive step of {\tt DMseg} is the permutation test, which entails calculation of differential methylation regression for genome-wide CpGs. Long clusters may require as many as 5000 permutations to get a high-resolution p-value. This is necessary for accurate assessment of the type I error rate for candidate DMR findings, the distictive feature of {\tt DMseg}. Extending this algorithm to whole-genome bisulphite sequencing data and accounting for variability of read depth is feasible, as have been done by {\tt DMRcate} \cite{Peters2021}.

\bibliography{DMseg_Dai}

\end{document}